\documentclass[aps,prl,reprint,groupedaddress]{revtex4-2}
\usepackage{gensymb}
\usepackage{graphicx}
\usepackage{amsmath}
\usepackage[utf8]{inputenc}
\usepackage[T1]{fontenc}
\begin{document}

% Use the \preprint command to place your local institutional report
% number in the upper righthand corner of the title page in preprint mode.
% Multiple \preprint commands are allowed.
% Use the 'preprintnumbers' class option to override journal defaults
% to display numbers if necessary
%\preprint{}

%Title of paper
\title{Design Rules for Interconnects Based on Graphene Nanoribbon Junctions}

% repeat the \author .. \affiliation  etc. as needed
% \email, \thanks, \homepage, \altaffiliation all apply to the current
% author. Explanatory text should go in the []'s, actual e-mail
% address or url should go in the {}'s for \email and \homepage.
% Please use the appropriate macro foreach each type of information

% \affiliation command applies to all authors since the last
% \affiliation command. The \affiliation command should follow the
% other information
% \affiliation can be followed by \email, \homepage, \thanks as well.
\author{Kristi\=ans \v Cer\c nevi\v cs}
\author{Oleg V. Yazyev}
%\email[]{Your e-mail address}
%\homepage[]{Your web page}
%\thanks{}
%\altaffiliation{}
\affiliation{Institute of Physics, Ecole Polytechnique F\'ed\'erale de Lausanne (EPFL), CH-1015, Switzerland}

%Collaboration name if desired (requires use of superscriptaddress
%option in \documentclass). \noaffiliation is required (may also be
%used with the \author command).
%\collaboration can be followed by \email, \homepage, \thanks as well.
%\collaboration{}
%\noaffiliation

\date{\today}

%K:Abstract should be 150 words for Nature Electronics, the current one is 218. Is this something they enforce?
\begin{abstract}
%Graphene nanoribbons (GNRs) are considered promising candidates for the next-generation electronic devices due to the broad range of properties that can be achieved depending on their width, crystallographic orientation and edge termination. Control over these characteristics is achieved using the bottom-up chemical self-assembly thus paving the way towards all-graphene integrated circuits of ultimate nanoscale dimensions. Junctions connecting two or more GNRs are essential components of such circuits, hence deep understanding of their electrical characteristics is critical. 
Graphene nanoribbons (GNRs) produced by means of bottom-up chemical self-assembly are considered promising candidates for the next-generation nanoelectronic devices.
We address the electronic transport properties of angled two-terminal GNR junctions, which are inevitable in the interconnects in graphene-based integrated circuits. We construct a library of over 400000 distinct configurations of 60$\degree$ and 120$\degree$ junctions connecting armchair GNRs of different widths. Numerical calculations combining the tight-binding approximation and the Green's function formalism allow identifying numerous junctions with conductance close to the limit defined by the GNR leads. Further analysis reveals underlying structure-property relationships with crucial roles played by the bipartite symmetry of graphene lattice and the presence of resonant states localized at the junction. In particular, we discover and explain the phenomenon of binary conductance in 120$\degree$ junctions connecting metallic GNR leads that guarantees maximum possible conductance. Overall, our study defines the guidelines for engineering GNR junctions with desired electrical properties.
\end{abstract}

% insert suggested keywords - APS authors don't need to do this
%\keywords{}

%\maketitle must follow title, authors, abstract, and keywords
\maketitle

% body of paper here - Use proper section commands
% References should be done using the \cite, \ref, and \label commands

%\section{Introduction}

Since the pioneering work on isolation and characterization of graphene by Novoselov \textit{et al.} \cite{Novo04}, this two-dimensional material and derived nanostructures have been attracting ever-growing interest. Variety of novel physical properties such as massless Dirac fermion quasiparticles \cite{Novo05a}, very high charge-carrier mobilities \cite{Bolo08} 
and more importantly, ballistic transport over micrometers length-scales \cite{Novo04}, has been revealed. Furthermore, graphene has emerged as a promising candidate for applications in nanoelectronics \cite{Geim07a,Geim09a,Ares07}, however its use in electric devices is limited due to the lack of band gap and thus poor switching capability \cite{Novo04}.
Among numerous approaches for opening a band gap in graphene \cite{Balo10,Ni08,Giov07}, geometric confinement \cite{Waka99a,Son06a,Baro06} in one-dimensional graphene nanoribbons (GNRs) being the prototypical systems has allured the largest interest in the field. 
However, the complex dependence of the electronic structure of GNRs on their structural characteristics \cite{Yazyev13}---width, crystallographic orientation and edge termination---renders top-down approaches unsuitable for achieving sufficient control over the band gap.

More recently, the bottom-up technique relying on on-surface self-assembly starting from predefined molecular precursors allowed synthesizing GNRs of desired structure with atomic precision \cite{Cai10a}. This approach made it possible to control the electronic and transport properties of GNRs by varying width \cite{Chen13,Kimo15}, edge structure \cite{Vo14,Liu15a,Pizz21b} and substitutional doping \cite{Clok15,Kawa15}. 
Same approach also allows producing more complex structures such as junctions joining multiple GNRs, a crucial ability for realizing functional  nanoelectronic integrated circuits based on graphene \cite{Ares07,Kang13}. More specifically, selected two-terminal junctions \cite{Blan12a,Ma19a,Li19b,Cern20c}, heterojunctions \cite{Bron18a,Nguy17,Mara16,Jaco17,Chen15a,Cai14a}, junctions exhibiting topological interface states \cite{Gron18,Rizz18} and even multiple-terminal junctions \cite{Cai10a}
synthesized by on-surface self-assembly have been reported. 

Interconnects joining individual electrical components are the most basic and fundamental building blocks of any integrated circuit. Linear interconnects based on one-dimensional GNRs allow for little freedom in designing such circuits, while to effectively accommodate a large number of components on a plane one would desire graphene interconnects that follow paths with ``turns'' akin to the conductive traces of common printed circuit boards. Each ``turn'' can be viewed as a two-terminal GNR junction, and the electronic transport properties of such a junction are expected to depend strongly on the details of its atomic structure. This is in contrast to macroscopic interconnects such as printed circuit board traces or ordinary wires, since bending a wire does not generally affect its resistance. While the electronic and transport properties of one-dimensional GNRs are largely understood by now, relations between the structure and electronic transport properties of GNR junctions remain to be explored.

In this work, we address the structure-property relations of two-terminal GNR junctions and formulate design rules for engineering interconnects with minimal scattering for graphene-based nanoelectronic circuits. To achieve this goal, we systematically characterize over 400000 unique
structures of GNR junctions connecting two identical armchair GNRs of different width at angles of 120 and 60 degrees. Extensive library of calculations performed using a tight-binding model and Green's function technique allows to
uncover several universal physical mechanisms underlying the electronic transport across GNR junctions and establish design rules necessary for engineering nanometer-width interconnects for graphene-based nanoelectronic circuits.  

\vskip 0.25cm
\noindent
{\bf Results and Discussion} \\
\noindent
Our work relies on the one-orbital nearest-neighbor tight-binding model Hamiltonian in conjunction with Green's function techniques for simulating the ballistic electronic transport properties (see Methods). This minimalist model allows for sufficiently accurate description of the discussed properties, yet it is simple enough for performing massive high-throughput screening of a large number of junction structures and rationalizing the numerical results in terms of analytical models. 

\begin{figure}[]
	\includegraphics[width=\columnwidth]{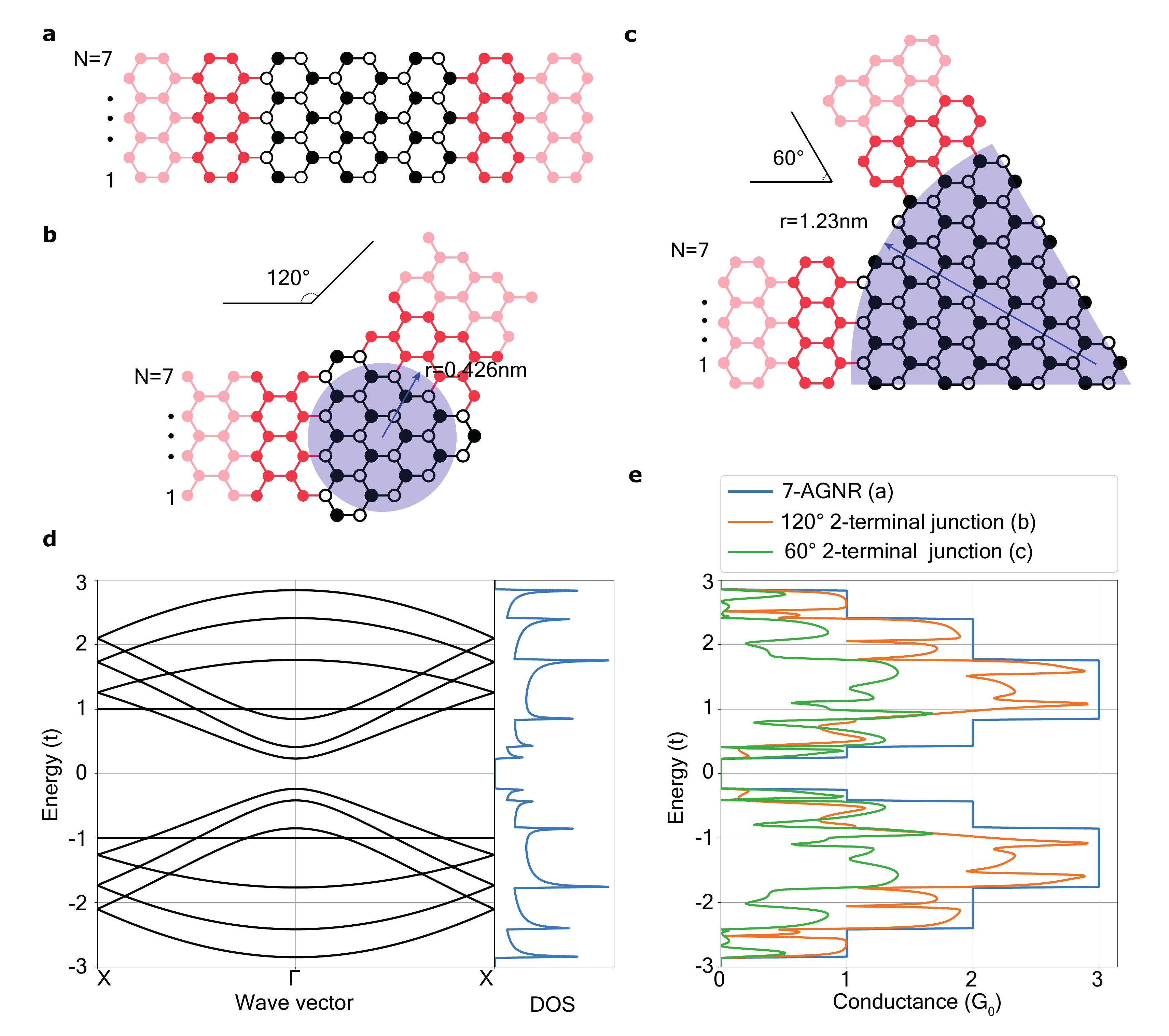}%
	\caption{GNR junctions and their electronic properties. (a) Atomic structure of 7-AGNR, with black and white atoms representing the two sublattices and red atoms showing the semi-infinite leads. (b) An example of 120$\degree$ two-terminal junction with 7-AGNR leads ($\mathit{IP_{C}}$). Area limiting the scattering region used in constructing various junction configurations is shown in blue. (c) An example of 60$\degree$ junction with 7-AGNR leads. (d) Band structure and DOS of 7-AGNR. (e) Conductance profiles of 7-AGNR lead (panel (a)), 120$\degree$ (panel (b)) and 60$\degree$ (panel (c)) junctions.\label{fig1}}
\end{figure}

{\it Construction of junction structures.---}Using this methodology we perform an exhaustive investigation of the electronic transport properties of 120$\degree$ and 60$\degree$ two-terminal GNR junctions connecting two armchair graphene nanoribbon (AGNR) leads of the same width. These junction angles, in addition to the trivial case of 180$\degree$, can be constructed without introducing any topological defects changing the crystallographic orientation of graphene lattice. 
We explored leads of different width defined by $3\leq N\leq 9$ carbon atoms, hence covering the semiconducting ($N=3p$, $N=3p+1$) and metallic ($N=3p+2$) AGNR families \cite{Naka96}.
Fig.\ \ref{fig1}(a) shows an example  $N = 7$ AGNR.

We then generate junction structures within the limits defined by a circular area as shown in Fig.\ \ref{fig1}(b) for the 120$\degree$ junctions of 7-AGNRs. The circle is defined by the intersection point of the lead axes such that there is no overlap between the leads.
Note, the position of the circle center leads to 3 distinct classes of 120$\degree$ junctions, which are denoted by lead intersection point $\mathit{IP_{I}}$, with $I=A,B$ or $C$ corresponding to sublattice A, sublattice B or the center of the hexagon [Supplementary Fig.~1]. The edge carbon atoms are assumed to be hydrogen terminated, but due to the absence of $\pi$-bonding hydrogen is not included in the one-orbital tight-binding model \cite{Hanc10}. Once the circular area is defined, we generate {\it all} possible junction configurations by removing atoms from it  and imposing the following restrictions: mirror symmetry of the junction and the presence of only three- and two-fold coordinated carbon atoms. For 60$\degree$  junctions, the circular area is defined in a similar way, except that the circle center is defined as the intersection of lines drawn along the GNR edges [Fig.\ \ref{fig1}(c)]. 
%The area is obtained once again by drawing a circle on the intersection point of the two leads, but is then sliced to align with the lead edges. Due to our imposed mirror symmetry, each 60$\degree$ 2-terminal junction has only one possible location of the circle's center. 

The properties of the generated junctions are then compared to those of pristine AGNR leads. Figure\ \ref{fig1}(d) shows the band structure and density of states (DOS), while Fig.\ \ref{fig1}(e) compares the conductance profiles of the three structures presented in Figs.\ \ref{fig1}(a)$-$(c). The results are presented in terms of the nearest-neighbor hopping integral $t=2.75$ eV \cite{Kund09a}. 
Quantized conduction can be observed for the pristine 7-AGNR, reflecting the number of sub-bands (transmission channels) at a particular energy $E$, with a maximum value $G=3G_{0}$ ($G_{0}=2e^2/h = 7.75 \times 10^{-5}$~S) when three channels are present and a minimum of $G=0$ in the band gap. The conductance of junctions is bound from above by the lead conductance and is no longer quantized 
due to scattering. In general, GNR junctions show complex conductance profiles with pronounced differences between each other. For example, the 60$\degree$ junction exhibits conductance close to the quantized conductance of the lead near the band edge ($E=0.3t$), but the conductance of 120$\degree$ junction is suppressed at the same energy.

The objective of designing optimal graphene-based interconnects consists in minimizing the amount of scattering at each ``turn'', \textit{i.e.} achieving conductance close to the limit defined by the conductance of the GNR leads. To quantify the junction conductance with respect to the ideal lead, we introduce a descriptor $\tau$ \cite{Pizz21a}
\begin{equation*}
\tau= \frac{\int_{E_0}^{E_0+\delta E}G_{j}(E)dE}{\int_{E_0}^{E_0+\delta E}G_{l}(E)dE} ,
\end{equation*}
which estimates the preserved conductance in a narrow energy window $\delta E=0.037t=0.1$~eV with $E_0$ being the conduction band minimum and the Fermi level for the semiconducting and metallic GNR leads, respectively. The chosen energy range roughly represents the expected operating conditions of graphene devices. $G_{j}(E)$ and $G_{l}(E)$ are the conductances of the junction and the ideal lead, respectively. From here on, we will be referring to well-conducting junctions if $\tau>0.9$.
 
\begin{figure}[]
	\includegraphics[width=\columnwidth]{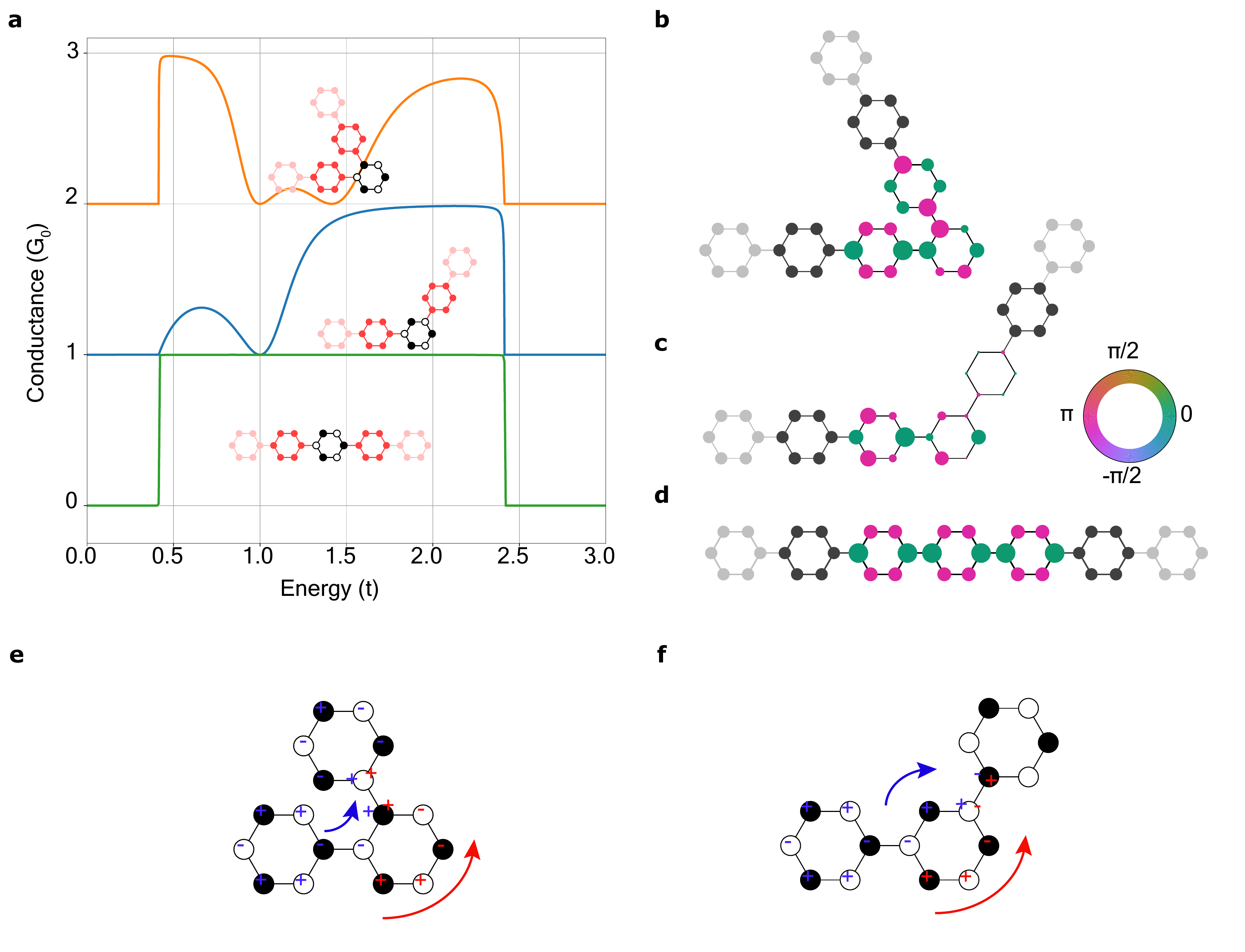}%
	\caption{Quantum interference in 3-AGNR junctions. (a) Conductance plots of \textit{ortho}-, \textit{meta}- and \textit{para}-attached 3-AGNR leads with the atomic structures displayed. The conductance profiles are offset with respect to each other by $G_0$ for clarity.
	The wave function inside the scattering region of (b)\textit{ortho}-, (c)\textit{meta}- and (d)\textit{para}-attached 3-AGNR junctions at $E=0.4t$. Schematic drawing of the two pathways (blue and red) and the phase in (e)\textit{ortho}- and (f)\textit{meta}-attached 3-AGNR junctions.} \label{fig2}
\end{figure}

{\it 3-AGNR two-terminal junctions.--}
We start our discussion of structure-property relationships by considering the simplest case of 3-AGNR junctions. Three possible configurations of the scattering region correspond to a single benzene ring connected to the leads in  \textit{ortho}-, \textit{meta}- or \textit{para}-positions, using chemistry notations, to obtain 60$\degree$, 120$\degree$ and 180$\degree$ junctions, respectively. The latter case is the ideal 3-AGNR, and hence no scattering is possible.  Fig. \ref{fig2}(a) shows the atomic structures of the junctions and displays the corresponding conductance spectra. We show only positive energies $0\leq E \leq 3t$ since at negative energies the results are identical due to the electron-hole symmetry inherent to our tight-binding model.

In Fig.\ \ref{fig2}(a), the 60$\degree$ junction demonstrates conductance very close to that of the perfect 3-AGNR lead ($\tau=0.99$) at the band edge ($E=0.4t$), whereas for the 120$\degree$  junction the conductance is essentially zero at this energy. One needs to note that seemingly small difference in the attachment points gives rise to distinct transport properties. Further differences between the three configurations can be observed in the scattering center wave functions at $E=0.4t$ plotted in Figs.\ \ref{fig2}(b),(c) and (d). When the leads are attached in the \textit{ortho}- position, the wavefunction is delocalized over all atoms in the scattering center but the phase is reversed upon transmission [Fig.\ \ref{fig2}(b)]. In contrast, the wave function is localized only on one of the sublattices in the central region of the 120$\degree$ junction and vanishes completely as it approaches the other lead [Fig.\ \ref{fig2}(c)], hence manifesting in nearly zero conductance.  

The aforementioned observations can be rationalized in terms of the quantum interference (QI) phenomenon. In Fig.\ \ref{fig2}(e), we show a schematic representation of the 60$\degree$ junction and graphically represent the phase of the incoming wave function as $+$ or $-$ as a continuation of the wavefunctions of the perfect 3-AGNR lead (Fig.\ \ref{fig2}(d)). We note the two inequivalent pathways colored in blue and red result in constructive QI and explain the phase reversal as observed in Fig.\ \ref{fig2}(b). Similarly, for the 120$\degree$ junction [Fig. \ref{fig2}(f)] the two pathways result in destructive interference. This simple model correctly predicts the wave functions in the scattering region [Figs.\ \ref{fig2}(b) and (c)] and reaches an excellent agreement with a graphical model designed for visual inspection of the connectivity in $\pi$-conjugated systems \cite{Zhao17a}.

Although QI has been observed to influence electronic transport properties in organic molecules \cite{Saut88,Li19a}, only recently its effect has been investigated for more complex graphene nanostructures. Calogero \textit{et al. }\cite{Calo19a,Calo19b} showed that the electronic transport in nanoporous graphene can be controlled  through \textit{para}- and \textit{meta}-bridges. However, we observe that in GNR junctions QI effects are pronounced only in cases where the scattering center is attached to the lead through a single bond. For wider AGNR junctions involving multiple covalent bonds to leads it becomes difficult to interpret the results in terms of QI, and thus systematic numerical characterization of the electronic transport properties needs to be performed. 

\begin{figure*}[]
	\includegraphics[width=2\columnwidth]{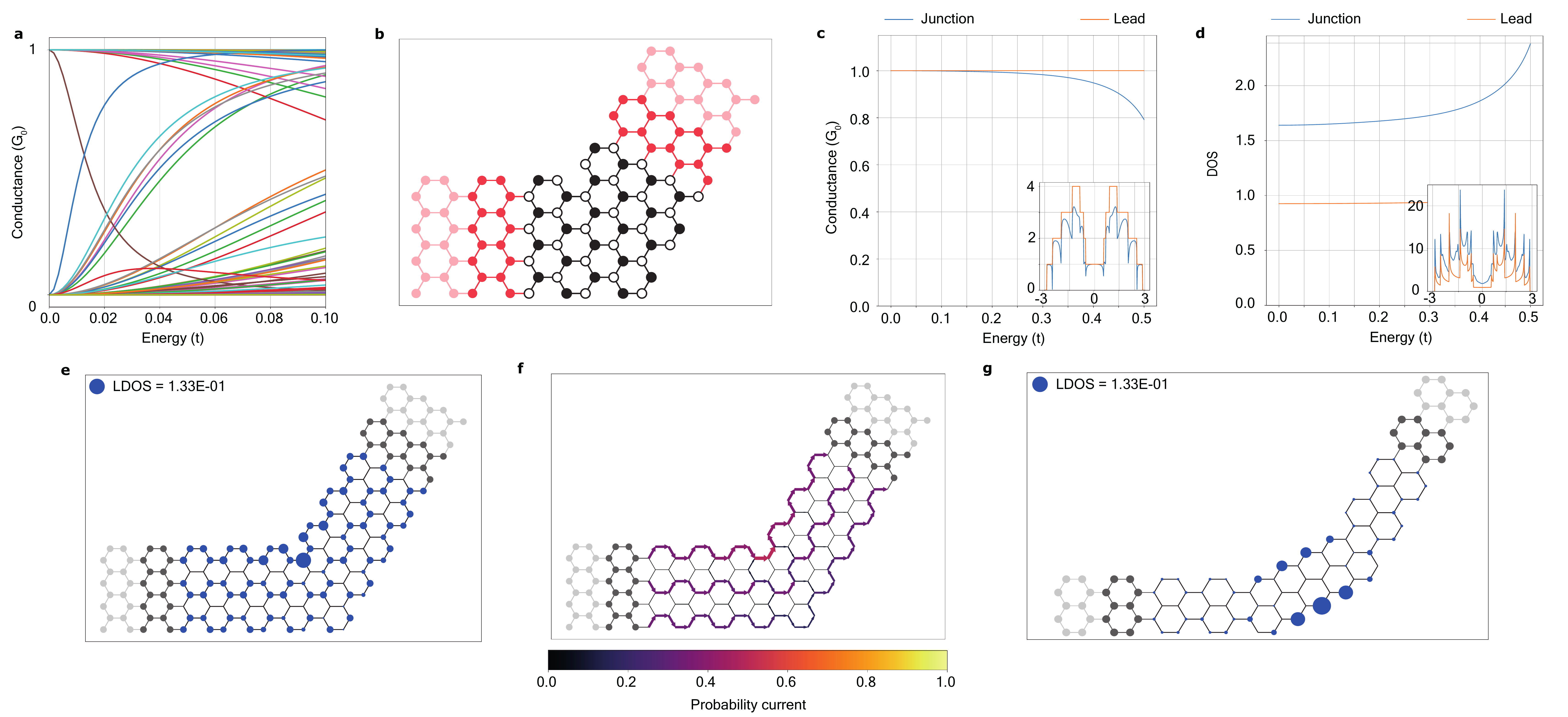}%
	\caption{Electronic transport across metallic 120$\degree$ junctions. (a) Conductance profiles of 150 representative 120$\degree$ 8-AGNR junctions in $0\leq E \leq0.1t$ energy range. (b) Atomic structure of a selected 120$\degree$ 8-AGNR junction with $\tau=1$ and (c) its conductance profile in the $0\leq E \leq0.1t$ energy range. The full conductance profile is shown in the inset. (d) DOS of the junction in $0\leq E \leq0.1t$ energy range and the its full profile (inset). (e) Local DOS and (f) local current across the junction at $E=0$. (g) Local DOS of a metallic 120$\degree$ 5-AGNR junction with zigzag edges at $E=0$.}
	\label{fig3}
\end{figure*}

{\it Systematic screening of electronic transport across junctions.--}
The structures of $N$-AGNR junctions with lead widths $4\leq N\leq 9$ were systematically constructed resulting in total of 438187 unique configurations. Regardless of the lead width and junction angle, it was always possible to identify configurations with $\tau \geq0.9$. The entire database of investigated junctions ordered in terms of $\tau$ is provided in Supplementary Tables 1--36. Below, we focus on representative junctions that demonstrate conductances close to those of ideal leads and help establishing the underlying structure-property relationships.

\begin{figure}[]
	\includegraphics[width=\columnwidth]{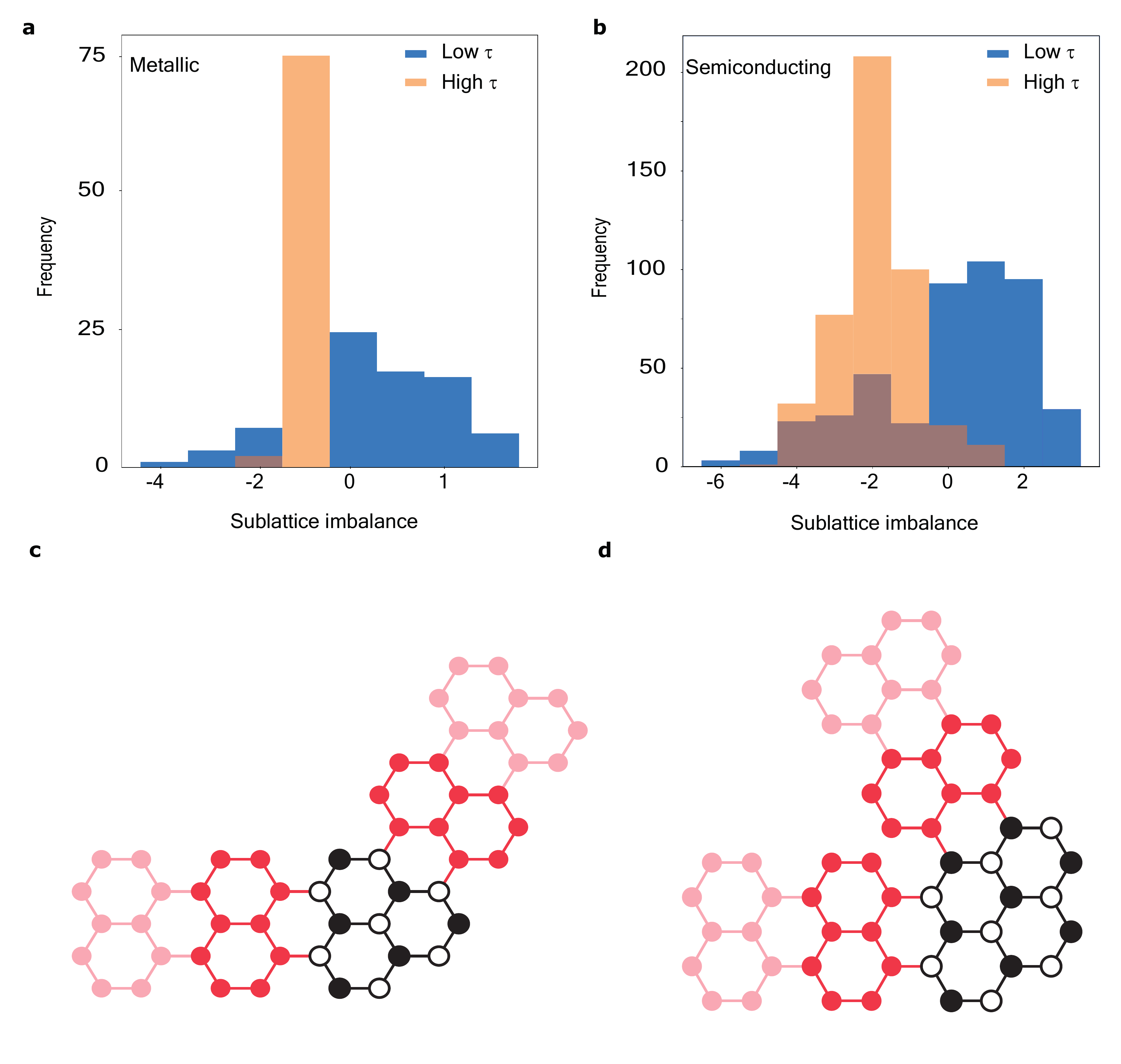}%
	\caption{Effect of sublattice imbalance on the transmission. (a) Distribution of sublattice imbalance $\delta N$ for 75 highest and 75 lowest (out of 712) values of $\tau$ among the considered 120$\degree$ 5-AGNR ($\mathit{IP_{C}}$) junctions. (b) Distribution of sublattice imbalance $\delta N$ for 450 highest and 450 lowest (out of 6817) $\tau$ values among the 120$\degree$ 7-AGNR ($\mathit{IP_{B}}$) junctions. Examples of (c) 120$\degree$ and (d) 60$\degree$ junctions with sublattices A and B shown in black and white, respectively.} 
	\label{fig4}
\end{figure}

{\it 120$\degree$ junctions.--}
We first discuss the 120$\degree$ junctions pointing a striking transport phenomenon observed for the metallic AGNR leads ($N =5,8$). Figure\ \ref{fig3}(a) shows conductance profiles of 150 randomly selected 8-AGNR junctions in the $0 \leq E \leq 0.10t$ energy range. The conductance at $E=0$ is binary taking only values of $G=G_{0}$ or $G=0$, and this effect is not observed for the 60$\degree$ junctions. We also note that both the 60$\degree$ and 120$\degree$ junctions with semiconducting leads do not show this effect at the band edge. In Fig.\ \ref{fig3}(b), we show a 120$\degree$ 8-AGNR junction that features $G=G_{0}$ at $E=0$ with practically no backscattering up to the energies as high as $E = 0.3t$ ($\sim 1$~eV) [Fig.\ \ref{fig3}(c)]. For this particular junction, there is no large variation in the DOS of the junction compared to the 8-AGNR lead [Fig.\ \ref{fig3}(d)], and the local density of states (LDOS) is rather delocalized in the scattering region at $E=0$ [Fig.\ \ref{fig3}(e)]. Our observation is in agreement with the finding that a delocalized transmission eigenstate leads to conductance enhancement \cite{Xia20a}. We also show that such delocalization leads to homogeneous local current paths through the junction [Fig. \ref{fig3}(f)] closely resembling local current in the lead. 
Remarkably, we also find that such perfectly transmitting 120$\degree$ junctions can demonstrate a DOS peak at $E=0$ and localized states associated with the zigzag edge segments [Fig.\ \ref{fig3}(g)]. We note that the electronic characteristics of this junction indicate a weakly coupled localized state leading to the Fano resonance \cite{Miro10}, as previously observed for T-shaped junctions \cite{Kong10} and quantum dots \cite{Xion11a}. We also find metallic 120$\degree$ junctions [see e.g. Supplementary Fig.~2] with perfect transmission that combine different types of edges--armchair, zigzag and mixed--thus concluding that the edge geometry in the scattering center does not play important role in determining the transport properties. However, we note that all metallic 120$\degree$ junctions with $\tau>0.9$ are characterized by non-zero sublattice imbalance $\delta N = N_A - N_B$, with $N_A$ and $N_B$ being the number of atoms in the scattering area belonging to sublattices A and B, respectively.
Fig.\ \ref{fig4}(a) shows a histogram of sublattice imbalance $\delta N$ for 75 highest and 75 lowest transmitting metallic (5-AGNR, $\mathit{IP_{A}}$) junctions. All junctions with high transmission have the same sublattice imbalance $\delta N=-1$ with an exception of two structures characterized by $\delta N=-2$. This deviation can be rationalized in terms of LDOS that is only localized in a sub-region with $\delta N=-1$ [see Supplementary Fig.\ 3 for discussion]. Similar behavior with decoupled localized states is observed in GNRs with edge functionalized molecules \cite{Cern20a}. We have also observed that for metallic 8-AGNR junctions the sublattice imbalance $\delta N=-3$ is associated with high $\tau$ value, whereas other sublattice differences $\delta N$ produce $G=0$ at $E=0$. The increase in $\delta N$ arises due to the two extra missing atoms at the interface between the scattering region and the lead for AGNRs with even number of carbon atoms across the width.
 
Therefore, we conclude that high conductance of 120$\degree$ junctions with metallic AGNR leads is associated with negative ($\delta N<0$) and odd ($\delta N \mod   2 \neq 0$) sublattice imbalance. 
In order to rationalize the observed binary conductance at $E=0$, we first note that in 120$\degree$ junctions the semi-infinite leads are connected to the same sublattice of the scattering region [Fig.\ \ref{fig4}(c)], while the opposite is true for the 60$\degree$ [Fig.\ \ref{fig4}(d)] and 180$\degree$ two-terminal junctions.
Second, in the nearest-neighbor TB model the number of zero-energy states is equal or larger the absolute value of $\delta N$. The wavefunction of a such zero-energy states is localized on the majority sublattice \cite{Wang09,Pere08}. It has been further shown that zero-energy states due to the sublattice imbalance facilitate the transmission of charge carriers \cite{Kihi17a}. Thus, we  conclude that the resonant zero-energy states are involved in the electronic transport resulting in maximum conductance at $E=0$ for metallic 120$\degree$ junctions. Depending on the coupling strength between the leads and the scattering centre, the DOS peak width of the zero-energy states can be significantly broadened and hence facilitate the conductance over a wide energy range around $E=0$. We show the conductance through a zero-energy state depending on the coupling strength with the leads in Supplementary Fig.\ 4. In contrast, metallic 120$\degree$ junctions with no sublattice imbalance ($\delta N=0$) or with a zero-energy state localized on the sublattice not connected to the leads ($\delta N>0$) exhibit $G=0$ at $E=0$. In these cases we observe that the trace of the matrix product in Eqn.~(\ref{eq.trans}) is zero due to diagonal elements having a pair with an opposite sign that stems from the symmetry of the device Hamiltonian and the fact that the self-energies $\sigma_{L(R)}$ in Eqn.~(\ref{eq:green}) affect only one sublattice. To support this finding, we present analytical derivation of the binary conductance phenomenon for metallic 120$\degree$ junctions in the Supplementary material. 

Figure\ \ref{fig4}(b) presents the histogram of sublattice imbalance $\delta N$ for 900 semiconducting 120$\degree$ junctions (7-AGNR, $\mathit{IP_{B}}$) with highest and lowest conductance. Interestingly, the junctions with highest $\tau$ have predominantly sublattice imbalance of $\delta N=-2$, while $\delta N\geq0$ is associated with low $\tau$ values. The results seem to indicate that the zero-energy states due to the sublattice imbalance still have a strong effect on electronic transport at the band edge.
We note, however, that this behavior is difficult to quantify and further investigation is required. Finally, another common features observed for $\tau>0.9$ semiconducting 120$\degree$ junctions is a narrow DOS peak at the band edge, resembling the DOS of the lead. The broadening of the DOS peak, on the other hand, leads to increased ``smoothening'' of the conductance profile and hence lower $\tau$ values. In extreme cases, we observe very small values of DOS near the band edge thus leading to $\tau=0$ [Supplementary Fig.\ 5]. 

\begin{figure*}[]
	\includegraphics[width=2\columnwidth]{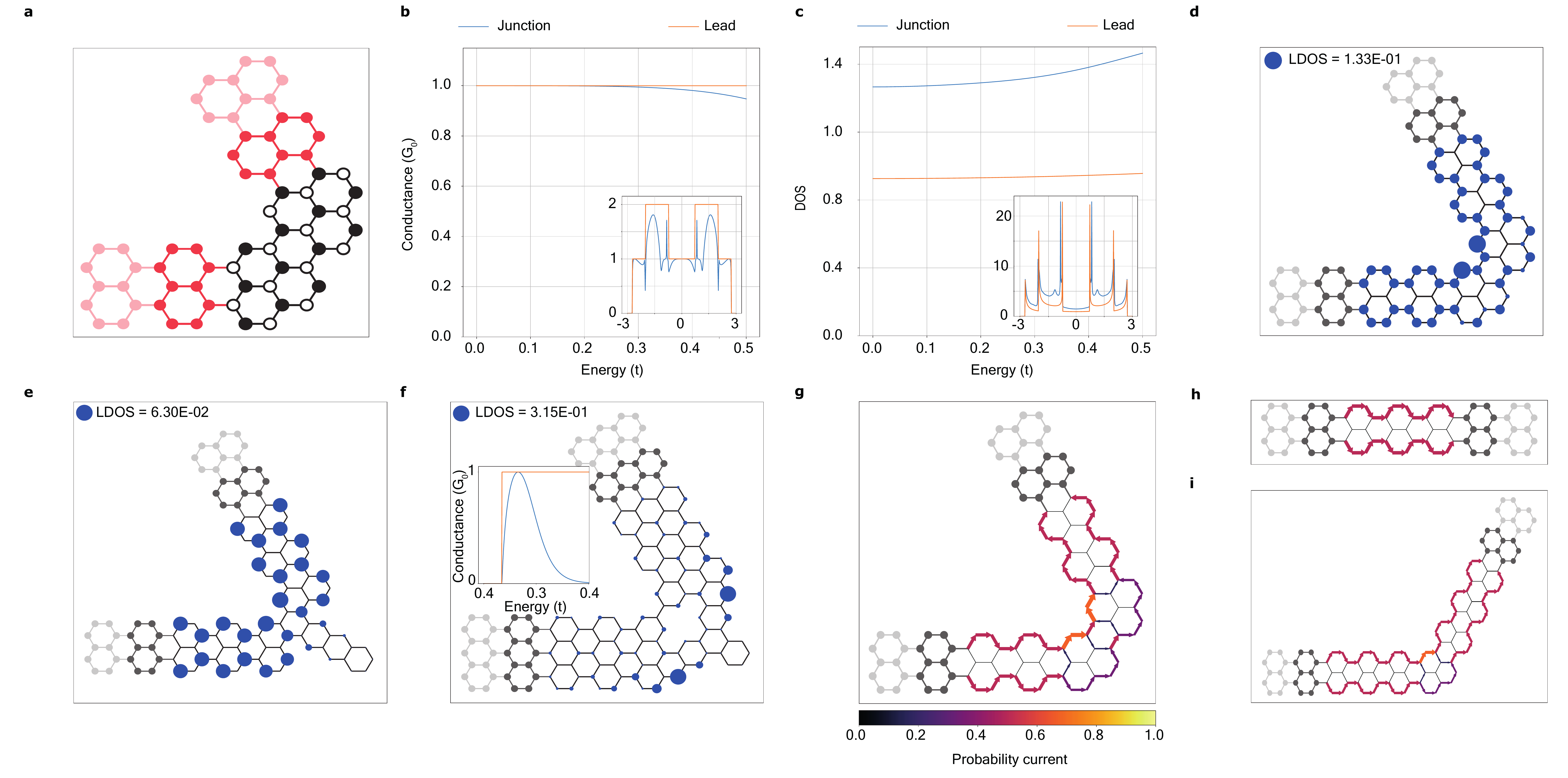}%
	\caption{Electronic transport in 60$\degree$ GNR junctions. (a) Atomic structure of a metallic 60$\degree$ 5-AGNR junction with $\tau=1$. (b) Conductance profile and (c) DOS of the junction in the $0 \leq E \leq 0.5t$ energy range and the full profiles (insets). (d) LDOS of the junction at $E=0$. (e) LDOS at $E=0$ of a selected metallic  60$\degree$ 5-AGNR junction characterized by $\tau=0$ at $E=0$. (f) LDOS at $E=0$ of a selected semiconducting 60$\degree$ 7-AGNR junction with $\tau=0.83$. (g) Probability current  at $E=0$ in the scattering region of the junction shown in panel (a). (h) Probability current at $E=0$ in the pristine 5-AGNR lead. (i) Probability current in a 120$\degree$ 5-AGNR junction at $E=0$ that represents half the 60$\degree$ 5-AGNR junction in panel (a).}
	\label{fig5}
\end{figure*}

{\it 60$\degree$ junctions.--}
As far as 60$\degree$ junctions are concerned, the leads are connected to the complementary sublattices of the scattering region and hence no binary conductance at $E=0$ is observed. Furthermore, all considered 60$\degree$ junctions have no sublattice imbalance since only symmetric configurations are investigated in our work.
Nevertheless, several common properties are observed for the 60$\degree$ and 120$\degree$ junctions with high $\tau$ values. In Fig.\ \ref{fig5}(a) we show a representative 5-AGNR 60$\degree$ junction with $\tau=1$. The conductance profile of the junction practically matches the one of the lead in the $0\leq E\leq 0.5t$ energy range [Fig.\ \ref{fig5}(c)]. Moreover, DOS in Fig.\ \ref{fig5}(c) is nearly constant, similar to the selected metallic 120$\degree$ junction in Fig.~\ref{fig3}(d), and the LDOS at $E=0$ reveals that the electron density is delocalised over the scattering region [Fig.\ \ref{fig5}(e)]. The effect of the localized states on conductance of 60$\degree$ junctions is however different. While localized states due to sublattice imbalance facilitate transmission across 120$\degree$ junctions, increased localization due to the presence of zigzag edges often leads to low $\tau$ values in metallic 60$\degree$ junctions.
In extreme cases of $\tau=0$ we notice that the electron density is localized only on one of the sublattices [Fig.\ \ref{fig5}(e)]. However, rare exceptions are found among semiconducting 60$\degree$ junctions where Fano resonances are observed near the band edge leading to high  values of $\tau$ [Fig.\ \ref{fig5}(f)]. These Fano resonances occur when localized states in the  scattering center hybridize with the continuum of states in the lead \cite{Miro10}. Overall, we observe that in 60$\degree$ junctions with $\tau>0.9$ local density of states resembles that in the leads and localization leads to lower transmission.  

Analyzing local current distributions further helps establishing the design rules of GNR junctions. It can be seen that the current is mainly carried by well defined channels, in the case of 5-AGNR leads localized at the edges, and is preserved in the scattering region [Fig.\ \ref{fig5}(d)]. Hence, comparing the probability current in the scattering center to the one of the pristine leads [Fig.\ \ref{fig5}(h)] gives an indication about the junction conductance. Strikingly, deviations such as the one observed in Fig.\ \ref{fig5}(g), where the local current on the inside edge is around 2.5 times larger than the current on the outside edge (matching the deviation of LDOS in Fig.\ \ref{fig5}(d)) can still lead to equal local current on both edges in the other lead. In general, we observe that junctions with preserved AGNR edges or even parts of scattering center resembling pristine AGNR geometry, usually display high $\tau$ values. This further supports the argument against inclusion of zigzag edges in the junctions as concluded from the LDOS analysis. Our observations agree with the work of Verges \textit{et al.} \cite{Verg18a}, where asymmetry and deviation from the lead geometry were found to decrease the conductance of the junctions. 

Finally, we show that 60$\degree$ junction with $\tau=1$ [Fig.\ \ref{fig5}(a)] can be viewed as two 120$\degree$ junctions with $\tau=1$ connected in series. As the local current leaving the junction in Fig.\ \ref{fig5}(i) matches the incoming current, the addition of a second junction with the same structure leads to a 60$\degree$ ``turn'' characterized by a high conductance. We also note that combining two equivalent 120$\degree$ junctions removes the sublattice imbalance and result in having both sublattices attached to the leads, thus eliminating the observed binary conductance phenomenon at $E=0$.

\vskip 0.25cm
\noindent
{\bf Conclusions} \\
In summary, we performed a systematic and exhaustive exploration of 120$\degree$ and 60$\degree$ two-terminal junctions in armchair graphene nanoribbons by means of numerical calculations combining tight-binding approximation and Green's function formalism. Our calculations show that irrespective of the lead width and junction angle it is always possible to construct junctions with minimal electron scattering, which could be used as optimal interconnects in all-graphene nanocircuits. 
Furthermore, having analyzed 438187 unique configurations we propose clear design rules to control the electronic transport properties of such junctions. 
In particular, we discovered digital, either full on or full off, conductance of metallic 120$\degree$ junctions, that is governed by sublattice imbalance. In contrast, for 60$\degree$-angled junctions highest conductance is achieved when armchair-type edge structure was preserved in the scattering center. %For the best structures, the scattering centers displayed very close similarities in LDOS and local current when compared to pristine leads. Curiously, we also found that two 120$\degree$ 2-terminal junctions with high conductance can be combined together to obtain 60$\degree$ 2-terminal junction without loosing the favorable electronic transport properties.

Finally, we present a complete library of results in Supplementary tables 1--36 \cite{Cern20d}, with junctions categorized by the width, angle and class. We also provide an easy-to-use online tool \texttt{TBETA} \cite{Cern20b} that allows reproducing all results presented in our work. Both the supplementary tables and the \texttt{TBETA} applications, which allows to construct arbitrary GNR junctions and calculate their electronic and  transport properties,  are publicly available on the Materials Cloud portal (https://www.materialscloud.org) \cite{Tali20}.
Overall, our results establish design rules necessary for engineering GNR junction with desired electrical characteristics.

\vskip 0.25cm
\noindent
{\bf Methods} \\
%\noindent
We employ a tight-binding (TB) model with one p$_{z}$ orbital per atom and only nearest-neighbour hopping integrals that has been shown to provide reasonably accurate description of the electronic structure of graphene \cite{Kund09a} and GNRs \cite{Hanc10} near the Fermi level. Importantly, this simple model is computationally inexpensive. We have extensively tested several TB models in our previous work \cite{Pizz21a} and provide further comparison of the nearest-neighbour TB model with DFT calculations in Supplementary Figure~6. The Hamiltonian is expressed as
\begin{equation}
H_{D}=\sum_{i}\epsilon_{i}c_{i}^{\dagger}c_{i}-t\sum_{i,j}(c_{i}^{\dagger}c_{j}+h.c.),
\label{eq:hamiltonian}
\end{equation} 
where $\epsilon_{i}$ is the on-site energy, $t$ is the nearest-neighbour hopping integral, $c_{i}^{\dagger}(c_{i})$ creates (annihilates) an electron on site $i$, while the sum $(i,j)$ is restricted to nearest-neighbour atoms. We set on-site energies $\epsilon_{i}=0$~eV and a hopping integral $t=2.75$~eV in our calculations.

Transport properties are calculated from the non-equilibrium Green's function (NEGF)
\begin{equation}
G_{f}(E)=\left((E+i\eta) I-H_{D}-\Sigma_{L}(E)-\Sigma_{R}(E)\right)^{-1},
\label{eq:green}
\end{equation} 
where $\eta$ adds an infinitesimally small imaginary character to  energy $E$ to avoid singular matrices, $I$ is the identity matrix and $\Sigma_{L(R)}$ is the self-energy containing information about the left (right) semi-infinite lead. The self-energies are obtained self-consistently using
\begin{equation}
\Sigma_{L(R)}(E)=H_{1}^{\dagger}(EI-H_{0}-\Sigma_{L(R)}(E))^{-1}H_{1},
\label{eq:dyson}
\end{equation}
where $H_{0}$ is the Hamiltonian of the unit cell of the lead and $H_{1}$ is the coupling between the lead unit cells. The broadening function $\Gamma_{L(R)}$ due to the leads is calculated from the self-energies
\begin{equation}
\Gamma_{L(R)}(E)=i[\Sigma_{L(R)}(E)-\Sigma_{L(R)}(E)^{\dagger}].
\label{eq:gamma}
\end{equation}
Next, the transmission is obtained by taking the trace of the following product
\begin{equation}
T(E)=Tr[\Gamma_{L}G_{f}\Gamma_{R}G_{f}^{\dagger}].
\label{eq.trans}
\end{equation}
Finally, we use the transmission coefficient $T$ obtained above to express conductance $G$ in terms of conductance quantum $G_{0}$ using the Landauer formula \cite{Land57a}
\begin{equation}
G(E)=G_{0}T(E)=\dfrac{2e^{2}}{h}T(E).
\end{equation}

We invite the readers to use our web-based open-access application \texttt{TBETA} \cite{Cern20b} on the \texttt{Materials Cloud} portal (https://www.materialscloud.org) \cite{Tali20} that implements the described methodology and allows reproducing all our results. It is designed for easy-to-use construction of angled junction configurations and calculation of their electronic transport properties. Access to wider range of junctions can be achieved by manually selecting the scattering region, including leads of different width and shifting the position of leads in relation to each other. No coding knowledge is required to design a junction and run the calculations. Conductance and density of states (DOS) can be computed over selected energy windows, while local density of states (LDOS), local current and lead wave functions are computed at a selected energy. All results reported in our work and the web-based application use \texttt{Kwant} \cite{Grot14} as the engine for performing calculations.

\vskip 0.25cm
\noindent
{\bf Acknowledgments} \\
The authors are financially supported by the Swiss National Science Foundation (Grant No.\ 172534) and the NCCR MARVEL.

% Create the reference section using BibTeX:
%\newpage
% \bibliography{Angled_bib}

%\newpage\begin{figure*}[]
%	\includegraphics[width=8cm]{toc.png}%
%	\caption{TOC graphic}
%~\\
%TOC graphic
%	\label{toc}
%\end{figure*}

%apsrev4-2.bst 2019-01-14 (MD) hand-edited version of apsrev4-1.bst
%Control: key (0)
%Control: author (8) initials jnrlst
%Control: editor formatted (1) identically to author
%Control: production of article title (0) allowed
%Control: page (0) single
%Control: year (1) truncated
%Control: production of eprint (0) enabled
%

\end{document}